\documentclass[twocolumn,showpacs,preprintnumbers,amsmath,amssymb]{revtex4}
\usepackage{dcolumn}
\usepackage{bm}
\usepackage{graphicx}

\begin{document}
\title{Exactly pairing two-dimensional charged particles using a magnetic field}
\author{Wenhua Hai} \affiliation{
Department of physics, Hunan Normal University, Changsha 410081,
China \\ Department of physics, Jishou University, Jishou, Hunan,
China}

%\altaffiliation{Corresponding Author: W. Hai.}

\email{adcve@public.cs.hn.cn}
\begin{abstract}

It is demonstrated that a uniform magnetic field can exactly pair
the two-dimensional (2D) charged particles only for some quantized
magnetic intensity values. For the particle-pair consisting of two
like charges the Landau level of the center-of-mass motion is
multiple degenerate that implies the shell structure. However, any
particle-pair has only two non-degenerate relative levels, which
are associated with the diamagnetic and paramagnetic states
respectively. There exist a upper critical magnetic strength and a
lower critical magnetic length across which two like charges
cannot be paired. The theoretical results agree with the
experimental data on the sites and widths of the integral and
fractional quantum Hall plateaus in a CaAs-Al$_x$Ga$_{1-x}$As
heterojunction, that gives a new explanation for the quantum Hall
effect.

\end{abstract}


\maketitle

The motions of the charged particles and neutral atoms in a
uniform magnetic field have been long studied \cite{Zeeman,
Landau}, and some important experimental developments of this
topic were reported successively
\cite{Garton}-\cite{Klarenbosch2}. The corresponding theoretical
investigates reveal many interesting properties concerning the
quadratic Zeeman effects \cite{Zeeman, Edmonds}, Landau level
\cite{Landau, Holle}, classical and quantum chaos \cite{Friedrich,
Hasegawa}, and the integral and fractional quantum Hall effects
\cite{Eisenstein, Klitzing, Klitzing2, Tsui, Stormer, Pan,
Laughlin, Chang}. All of the works are correlated with the
fundamental system: a pair of charged particles interacting with a
uniform magnetic field, where both the Coulomb and harmonic
potentials work simultaneously. The previous works have shown that
the Coulomb-harmonic system is exactly solvable only for some
particular values of the harmonic oscillator frequency \cite{Taut,
Pino, Hai, Alberg}.

In this paper we shall extend the results to a 2D electronic gas
and seek its exact eigenstates and eigenenergies, where the
harmonic potential is supplied by the external magnetic field.
This extension will lead to some interesting new results that
demonstrate the possibility for using a magnetic field to pair the
charged particles, and suggests a new theory to confirm the
assertion on ``the fractional quantum Hall effect must result from
the condensation of the 2D electrons" \cite{Stormer}.

Quantum motion of the 2D charged-particle-pair in a uniform
magnetic field toward $z$ direction is dominated by the
stationary-state Schr\"{o}dinger equation \cite{Shi, Taut},
\begin{eqnarray}
\sum_{i=1}^2 \Big[&-& \frac{\hbar^2}{2 m_i} \bigtriangledown _i^2
 +\frac 1 2 m_i \omega_i^2 (x_i^2+y_i^2) \mp
\omega_i l_i+\frac {q_1 q_2}{\epsilon r}\Big]\Psi \nonumber \\
&=&E_T \Psi,
\end{eqnarray}
where $E_T$ is the total energy, $m_i$ and $q_i$ are the effective
mass and charge \cite{Klitzing2, Shi} of $i$-th particle, $x_i,
y_i$ and $\bigtriangledown _i^2$ are the coordinates and Laplace
operators of particle $i$, $r=\sqrt{(x_2-x_1)^2+(y_2-y_1)^2}$ is
the relative radial coordinate, $l_i=x_i p_{y_i}-y_i p_{x_i}$
denotes the $z$-component angular momentum of particle $i$,
$\omega_i=|q_i|B/(2m_i c)$ represents the Larmor \emph{cyclotron
frequency} with the light velocity $c$ and the strength $B$ of
magnetic field, and $\epsilon$ is the effective dielectric
constant of the background semiconductor \cite{Chang, Shi}. Here
and throughout the paper the above signs of $\pm$ and $\mp$ are
selected for the positive-charge system, and the below signs are
taken for the negative-charge system. In order to exactly solve
Eq. (1) we introduce the relative coordinate $\mathbf{r}$ and
center-of-mass coordinate $\mathbf{R}$ by
$\mathbf{r}=\mathbf{r}_2-\mathbf{r}_1=(x_2-x_1)
\mathbf{e}_x+(y_2-y_1)\mathbf{e}_y=x \mathbf{e}_x+y\mathbf{e}_y,
\mathbf{R}=(m_1\mathbf{r}_1+m_2\mathbf{r}_2)/M=X\mathbf{e}_x+Y\mathbf{e}_y$
with $M=m_1+m_2$ being the total mass, $\mathbf{e}_x$ and
$\mathbf{e}_y$ the unit vectors in $x$ and $y$ directions. Setting
$E_T=E_c+E$ and $\Psi=\psi^{(c)}(\mathbf{R})\psi(\mathbf{r})$,
under the \emph{separability condition} $q_1/m_1=q_2/m_2 \
(\omega_1=\omega_2=\omega)$, Eq. (1) is separated as the
center-of-mass motion equation
\begin{eqnarray}
\Big[-\frac{\hbar^2}{2 M} \bigtriangledown _R^2 +\frac 1 2 M
\omega^2 R^2 \mp \omega L_z\Big]\psi^{(c)}=E_c \psi^{(c)},
\end{eqnarray}
and the relative motion equation
\begin{eqnarray}
\Big[-\frac{\hbar^2}{2 \mu} \bigtriangledown _r^2
 +\frac 1 2 \mu \omega^2 r^2 \mp
\omega l_z+\frac {q_1 q_2}{\epsilon r}\Big]\psi =E \psi,
\end{eqnarray}
where $\mu=m_1m_2/M$ is the reduced mass, $L_z=X P_{y}-Y P_{x}$
and $l_z=x p_{y}-y p_{x}$ are the $z$-component angular momenta on
the center-of-mass and relative coordinate frames respectively.
Clearly, Eq. (3) includes the equation of a 2D hydrogen atom in a
uniform magnetic field \cite{Taut, Alberg}, which can be directly
derived from Eq. (1) by fixing the atomic nucleus and taking the
positive sign of $\mp$. Therefore, Eq. (3) can govern the particle
pairs consisting of the like charges (e.g. two electrons, two
holes or two protons) or the unlike charges (e.g. hydrogen atom or
hydrogenic donor-electron \cite{Zhu}). Noticing the Pauli's
exclusion principle, the two like particles should be in
\emph{opposite spin orientation}.

For two-electron (or two-hole) system with $m_1=m_2=m_e$ and
$|q_1|=|q_2|=e$, Eq. (2) becomes the normal equation of a single
charged particle with mass $M=2m_e$ and charge $|q|=2e$ in a
uniform magnetic field, which has the well known solution
$\psi^{(c)}=\psi^{(c)}_{n_cm_c} (R, \phi_c)= e^{i m_c
\phi_c-R^2/2}R^{|m_c|}F(-n_c,|m_c|+1,R^2), \ n_c=0,1,2,\cdots; \
m_c=0,\pm 1, \pm 2, \cdots$ for the center-of-mass energy (Landau
level) $E_c/(\hbar \omega)=2n_c+|m_c|\mp m_c+1=1,2,3,\cdots$. Here
$F(-n_c,|m_c|+1,R^2)$ denotes the confluent hypergeometric
function, $m_e$ is the rest mass of a free electron, and the
coordinate $R$ has been normalized by the magnetic length
$a_c=\sqrt{\hbar/(M\omega)}$. Obviously, $\psi^{(c)}_{n_cm_c}$
expresses the \emph{multiple degenerate states}, since the energy
$E_c/(\hbar \omega)=2n_c+1$ corresponds to the $M_{n_c}+1$ states
with $|m_c|\mp m_c=0$ for $m_c=0,\pm 1,\pm 2,\cdots,\pm M_{n_c}$.
The largest magnetic quantum number $M_{n_c}$ and principal
quantum number $n_c$ are limited by the mean square-radius (or
area) $\overline {R^2_{n_c}}_{m_c}=\langle
\psi^{(c)}_{n_cm_c}|R^2|\psi^{(c)}_{n_cm_c} \rangle /\langle
\psi^{(c)}_{n_cm_c}|\psi^{(c)}_{n_cm_c} \rangle \le R_0^2$ for a
2D circular system with radius $R_0$. Taking the lowest Landau
level $E_c=1(\hbar \omega)$ with $n_c=0$ as an example, this
equation gives $\overline {R^2_0}_{m_c}=(|m_c|+1)(a_c^2)\le
R_0^2$. Thus we have the \emph{degeneracy of the lowest Landau
level} $M_0+1=\max\{|m_c|\}+1= R_0^2/a_c^2=2\pi R_0^2 M\omega/h
=2\pi R_0^2 eB/(hc)=\Phi/\Phi_0$, where $\Phi_0=hc/(2e)$ is the
magnetic flux quantum of the electron-pairs and $\Phi=\pi R_0^2B$
denotes the magnetic flux through the sample, which equates the
integral times of $\Phi_0$.

We are interested in exactly solving the relative motion equation
(3) and using the solution to describe the properties of the
particle-pairs, including the application to the integral and
fractional quantum Hall effect. Adopting the normalized polar
coordinate $\rho =r/a_r, \ a_r=\sqrt{\hbar/(\mu \omega)}$ and the
separate wavefunction \cite{Taut}
\begin{eqnarray}
\psi (\rho, \phi)=A e^{i m \phi-\rho^2/2}\rho^{|m|}u(\rho), \ \ \
m=0, \pm 1, \cdots
\end{eqnarray}
with the normalization constant $A$, applying Eq. (4) to Eq. (3)
we arrive at the 1D dimensionless equation
$\frac{d^2u}{d\rho^2}+\Big(\frac{2|m|+1}{\rho}-2\rho\Big)\frac{du}{d\rho}+\Big[2\Big(\frac{E'}{\hbar
\omega}-|m|-1\Big)-\frac {\sigma}{\rho}\Big]u = 0,$ where the
energy has been changed to $E'=E\pm m\hbar \omega$. The
dimensionless constant $\sigma=2q_1q_2\sqrt{\mu/(\hbar^3\omega
\epsilon^2)}$ \emph{expresses the importance of the Coulomb
potential compared to the harmonic one}, since when $q_1=q_2=-e,
\mu=m_e/2$ and $\epsilon=1$ we have
$\sigma^2=\frac{2e^2/a_0}{\hbar \omega}$ with
$a_0=\hbar^2/(m_ee^2)$ being the Bohr radius. According to the
form of $u$ equation we expect the power-series solution
\cite{Hai} $u=\sum_{i=0}^nC_i\rho^i$ for $\ n=1,2,\cdots$ and
$C_i=$constant. Inserting the series into the $u$ equation yields
the algebraic equation $\sum_{i=0}^n\{(i^2+2i|m|)\rho^{i-2}+\sigma
\rho^{i-1}+2[E'/(\hbar \omega)-|m|-1-i]\rho^i\} C_i=0$. This
equation can be satisfied if and only if (iff) the constants $C_i,
\ \sigma$ and $E'$ obey the coefficient equations \cite{Hai}
$E'/(\hbar \omega)=n+|m|+1, \ (n-j)(n-j+2|m|)C_{n-j}-\sigma
C_{n-j-1}+2(j+2)C_{n-j-2}=0$ for $C_0=1, \ C_{-1}=C_{n+1}=0,  \
j=-1,0,1,\cdots,(n-1)$. By using a computer we have solved this
equation group for $n=1,2,\cdots,30$ and several simple solutions
are listed as the following:
\begin{eqnarray}
n=1,& \sigma^2=(\sigma_{1|m|}^{(l)})^2=2(2|m|+1),  \ \
C_1=\frac{\sigma_{1|m|}^{(l)}}{2|m|+1}; \ \   \nonumber \\ n=2,&
(\sigma_{2|m|}^{(l)})^2=4(4|m|+3),
C_1=\frac{\sigma_{2|m|}^{(l)}}{2|m|+1}=\frac{\sigma_{2|m|}^{(l)}}{2}C_2;
  \nonumber
\\ n=3,& (\sigma_{3|m|}^{(l)})^2=20(|m|+1)\pm 2\sqrt{64m^2+128|m|+73};  \nonumber
\\ n=4,&(\sigma_{4|m|}^{(l)})^2=50+40|m|\pm
6\sqrt{16m^2+40|m|+33},\
\end{eqnarray}
where we have set $l=1,2,\cdots,l_{max}, \ l_{max}=n$ for even $n$
and $l_{max}=(n+1)$ for odd $n$, which is the label of the
different solutions for a set of fixed quantum numbers $n$ and
$|m|$. In the cases $n=3,4$, any $C_i$ is a complicated function
of $|m|$, which is not shown in Eq. (5). When $n\ge 5$, the
computer cannot give $C_i$ and $\sigma$ as the explicit functions
of $|m|$, however, we can numerically calculate them for any given
$n$ and $|m|$. Noticing the above-mentioned relationships between
$\omega$ and $\sigma$, and $B$ and $\omega$, the quantized
$\sigma$ values imply the quantization of the cyclotron frequency
$\omega$, magnetic strength $B$ and energy $E$,
\begin{eqnarray}
\omega=\omega_{n|m|}^{(l)}=\frac{4q_1^2q_2^2\mu}{\hbar^3\epsilon^2(\sigma_{n|m|}^{(l)})^2},
\ B=B_{n|m|}^{(l)}=\frac{8|q_1|q_2^2\mu
m_1c}{\hbar^3\epsilon^2(\sigma_{n|m|}^{(l)})^2}, \nonumber \\
E_{nm}^{(l)}=E_{n|m|}^{'(l)}\mp m\hbar \omega_{n|m|}^{(l)}
=(n+|m|\mp m+1)\hbar \omega_{n|m|}^{(l)}. \
\end{eqnarray}
The dependence of $C_i$ and $\rho \ (a_r)$ on the quantum numbers
$n,|m|,l$ leads to the power-series $u=u_{n|m|}^{(l)}$ and the
relative wavefunction $\psi=\psi_{nm}^{(l)}$.

Given Eqs. (5), (6) and the label of the wavefunctions, we find
the following interesting properties:

\bf a\rm) Two like charges can be paired iff the magnetic strength
is quantized as in Eq. (6). The necessity and sufficiency of the
\emph{pairing condition} infer that the magnetic field fitted to
the electron-pairs will suppress the second pairing of four
electrons, since the latter does not satisfy the pairing condition
of the former. The size of the particle-pair in the states of
lower quantum numbers is in the order of the relative magnetic
length $a_r=a_{n|m|}^{(l)}=\sqrt{\hbar/(\mu
\omega_{n|m|}^{(l)})}$, that may approach the size of the lowly
excited hydrogen atom, when the magnetic field is strong enough.
For example, the magnetic strength value $B\sim 30$Tesla (T)
corresponds to the cyclotron frequency $\omega \sim 10^{13}$Hz and
magnetic length $a_r \sim 10^{-9}$m for $\mu=m_e/2$.

\bf b\rm) The cyclotron frequency and magnetic strength depend on
$\sigma^2$ rather than $\sigma$. This means that for a fixed
magnetic strength the \emph{electron-pairs $(\sigma>0)$ and
hydrogen atom $(\sigma<0)$ with same $|\sigma|$ have the same
level structure} determined by the cyclotron frequency. However,
the coefficients $C_i$ depends on $\sigma$ that results in the
different power-series solutions $u$ for the two different
systems. For instance, in the Coulomb repulsion case $(\sigma>0)$,
Eq. (5) implies the simplest power-series solution
$u=u_{1|0|}^{(1)}=(1+C_1 \rho)=(1+\sqrt{2}\rho)$, but in the
Coulomb attraction case $(\sigma<0)$, the simplest power-series
solution is $u=u_{1|0|}^{(2)}=(1-\sqrt{2}\rho)$. Substituting them
into Eq. (4) respectively, produce the mean radius $\overline
{r}_{1|0|}^{(1)}=\langle\psi_{1|0|}^{(1)}| r|\psi_{1|0|}^{(1)}
\rangle /\langle \psi_{1|0|}^{(1)}|\psi_{1|0|}^{(1)}\rangle
=1.15739(a_{1|0|}^{(1)})$ in the Coulomb repulsion state and
$\overline {r}_{1|0|}^{(2)}=1.45221(a_{1|0|}^{(1)})$ in the
Coulomb attraction state $\psi_{1|0|}^{(2)}$, the latter is
greater than the former. This is very important that the combined
action between the Coulomb repulsion and magnetic field make the
particle-pair of two like charges the tight-binding
quasi-particle.

\bf c\rm) The magnetic strength is correlated to $|m|$ but the
energy and wavefunction are correlated to $m$, that infers the
\emph{particle-pair under the magnetic field $B$ being a two
internal-level system with two non-degenerate relative states},
the paramagnetic state $(m>0)$ and diamagnetic state $(m<0)$. The
level difference between the two states reads $\Delta
E_{nm}^{(l)}=2m \hbar \omega_{n|m|}^{(l)}$. If the particle-pair
is composed of two positive charges, the sign $``-"$ in Eq. (6) is
taken that means the paramagnetic state being the ground state
$(|m|- m=0)$ and the diamagnetic state being the excitation one
$(|m|- m=2|m|)$. Conversely, the diamagnetic state of the
electronic particle-pair (hydrogen atom or two electrons) is the
ground state and the paramagnetic state is the excitation state.
Because the amplitude $|\psi|$ depends on $|m|$ rather than $m$ so
the paramagnetic and diamagnetic states have the same relative
probability distribution.

\bf d\rm) Differing from the Landau level of center-of-mass
motion, the \emph{larger quantum number $n$ corresponds to the
lower cyclotron frequency and relative energy for a given $m$}. As
$n$ tends to infinity, the frequency and energy tend to zero,
since the increase velocity of $\sigma^2$ is much greater than
that of the $n$, consequently, the relative magnetic length $a_r$
approaches infinity such that the particle-pair is ionized.
Similarly, for a given $n$ and in the ground state $(|m|\mp m=0)$
case, the larger $|m|$ corresponds to the lower $\omega$ and $E$,
and the infinite $|m|$ is associated with the zero energy and
unbound state, however, in the excitation state case $(|m|\mp
m=2|m|)$, the limit $\lim_{|m| \rightarrow \infty}E_{nm}^{(l)}$ is
equal to a constant, although $a_r$ still tends to infinity.

\bf e\rm) There are two kinds of the \emph{quantum transitions}
for the particle-pair. One is the transitions between the ground
and excitation states determined by a fixed magnetic field, which
can be operated by using a laser with the frequency $2|m|\omega$.
Because the amplitude $|\psi|$ only depends on $|m|$, the
transition between $m$ state and $-m$ state does not change the
relative probability distribution of the particle-pair, but varies
the direction of angular momentum. Another kind of transitions
occurs between the states with different magnetic strengthes.
Therefore, this kind of transitions can be controlled by adjusting
the magnetic strengthes from one value of $B_{n|m|}^{(l)}$ to
another. These quantum transitions are probably useful for
performing the \emph{quantum logic operations}.

\begin{table*}
\caption{\label{tab:table1}Comparison between the theoretical
$\nu_{n|m|}^{(l)}$ and experimental $\nu$}
\begin{ruledtabular}
\begin{tabular}{cccccccccccccccc}
 $\nu$&$\frac 2 7$&$\frac {5}{17}$&$\frac {1}{3}$&$\frac {4}{11}$&$\frac {5}{13}$
&$\frac {2}{5}$&$\frac {3}{5}$&$\frac {2}{3}$&$\frac {5}{7}$&$\frac {4}{5}$&1&$\frac {4}{3}$&$\frac {5}{3}$&2&3\\
\hline
 $_{n|m|}^{(l)}$&$_{5,1}^{(1)}$&$_{21,0}^{(1)}$&$_{1,2}^{(1)}$&$_{29,0}^{(1)}$&$_{3,2}^{(1)}$&$_{2,0}^{(1)}$&$_{1,4}^{(1)}$&$_{3,4}^{(1)}$&
 $_{5,4}^{(1)}$&$_{3,5}^{(1)}$&$_{1,7}^{(1)}$&$_{22,0}^{(1)}$&$_{1,12}^{(1)}$&$_{2,3}^{(1)}$&$_{1,22}^{(1)}$ \\
$\nu_{n|m|}^{(l)}$&$0.291$&$0.295$&$\frac 1 3$&$0.367$&$0.387$
&$\frac {2}{5}$&$\frac {3}{5}$&$0.659$&$0.715$&$0.794$&1&$1.324$&$\frac {5}{3}$&2&3\\
\end{tabular}
\end{ruledtabular}
\end{table*}

\bf f\rm) The parameter $|\sigma|$ has a minimal value
$|\sigma_{1,0}^{(1)}|$, which corresponds to the \emph{upper
critical magnetic field and lower critical magnetic length} across
which two like charges cannot be paired. For an electron-pair we
insert the parameters $q_1=q_2=-e, \ \mu=m_1/2=m_e/2, \
\epsilon=1$ into Eqs. (5) and (6), producing the largest magnetic
strength $B_{10}^{(1)}=2m_e^2e^3c/\hbar^3\approx 4.7\times 10^5$T,
which cannot be experimentally realized yet. The experimentally
allowable magnetic strength $B<10^2$T and the pairing condition of
two electrons require the parameter $\sigma^2$ being greater than
$10^3$. However, if one adopts the modulation-doped
CaAs-Al$_x$Ga$_{1-x}$As heterojunction and let the 2D electrons
exist in GaAs at the interface between GaAs and Al$_x$Ga$_{1-x}$As
\cite{Klitzing2, Laughlin, Shi}, the corresponding parameters
become \cite{Shi} $q_1=q_2=-e, \ \mu=m_1/2=0.067m_e/2, \ \epsilon
\approx 13$ such that Eq. (6) gives the largest magnetic strength
$B_{10}^{(1)}\approx 12.4842$T, such magnetic field is
experimentally realizable. The largest magnetic strength
$B_{10}^{(1)}$ determines the lower critical magnetic length
$a_{1|0|}^{(1)}$. By improving the quality of the
CaAs-Al$_x$Ga$_{1-x}$As interface, the effective mass $m_1$ and
dielectric constant $\epsilon$ can be adjusted in the regions $m_1
\ge 0.065m_e$ and $\epsilon\ge 1$ in a practical experiment
\cite{Kittel, Barmby}, therefore, we can obtain the upper critical
magnetic strength in a great region.

We now try to extend the above-mentioned results to a 2D
electronic gas. Because of the multiple degenerate states of the
center-of-mass motion and the two non-degenerate relative states,
we expect the existence of the \emph{shell structure} of the
electron-pairs. There may be $2(M_0+1)$ electron-pairs occupy the
lowest Landau level of the center-of-mass motion and two of them
labelled by $(m_c, m)$ and $(m_c, -m)$ have the mean square-radius
$\overline {R^2_0}_{m_c}$ for the fixed magnetic strength
$B_{n|m|}^{(l)}$. The multiple electron-pairs in the lowest Landau
level seem to behave like the condensed Bosons. According to the
superconductivity theory of the Cooper pairs \cite{Bardeen}, the
condensation of electrons is necessary for the resistance
vanishing approximately. Therefore, we expect the minima in the
resistivity appearing at the quantized magnetic strength values of
the pairing condition (6). This expectation has been proved by the
experimental data on the integral and fractional quantum Hall
effect \cite{Klitzing2, Tsui, Stormer, Pan}. In fact, rewriting
the pairing condition as
$B_{n|m|}^{(l)}=B_{1,7}^{(1)}/\nu_{n|m|}^{(l)}, \
\nu_{n|m|}^{(l)}=(\sigma_{n|m|}^{(l)})^2/30$ and comparing the
constant $\nu_{n|m|}^{(l)}$ with the filling factor $\nu$ at which
the diagonal part of the resistivity tensor vanishing or taking
minimum experimentally, good agreement is found in the
experimental accuracy, as in table 1. Here $B_{1,7}^{(1)}$ is the
magnetic strength at $\nu=1$, which is dependent of the sample
material and can be determined by the experiments \cite{Klitzing2,
Tsui, Stormer, Pan}. In table 1 we show the strict agreement
between the experimental $\nu$ and theoretical $\nu_{n|m|}^{(l)}$
for the experimentally strong minima of the resistivity and the
approximate agreement in $10^{-3}$ order for the experimentally
weak states, containing a lot of data disappearing in the table.
The small errors may be caused from the difference between the
finite temperature effect in the experiments  \cite{Tsui, Stormer}
and the zero temperature assumption in theory. It is worth noting
that the experimentally strong minima appear in $n=1, \ 2$ states
with lower relative energies. For example, the two neighbor states
$\nu_{21,0}^{(1)}\approx 5/17$ and $\nu_{1,2}^{(1)}=1/3$ have near
$\sigma$ and $\omega$ values, but Eq. (6) give their relative
energies as $E_{21,0}^{(1)}=22 \hbar \omega_{21,0}^{(1)}$ and
$E_{1,2}^{(1)}=\frac 1 3 E_{1,-2}^{(1)}\approx 2 \hbar
\omega_{21,0}^{(1)}$ respectively, the former is much greater than
the latter, so the $1/3$ state is much strong compared to the
$5/17$ state \cite{Pan}, since there are more electron-pairs to
occupy the lower energy states. The stronger state is associated
with the lower diagonal resistivity and wider Hall plateau in the
experiments.

In conclusion, we have revealed a new pairing mechanism of the 2D
charged particles, namely the combining interaction of the
quantized magnetic field and Coulomb potential governs the charged
particle-pairs. The exact pairing states and level structure are
found that show many new and important physical properties of the
system. Applying the results to the quantum Hall effect of 2D
electronic gas, we obtain the places and strengthes of the minimal
diagonal resistivity, which are in good agreements with the well
known experimental data.

At the end of the paper we must point out that the theoretical
results have hinted vaguely some connections between the
electronic pairing states and the superconductivity of high
critical temperature $(T_c)$, such as the 2D electron-pairs in the
quantum Hall system behaving like the Cooper pairs in the 2D
superconductive material, the property of the critical magnetic
strength exceeding the experimental limit being similar to that of
the upper critical field of the type II superconductor with high
$T_c$, and the lower critical magnetic length to be in the order
of the electronic correlation length of the high $T_c$
YBa$_2$Cu$_3$O$_{7-\delta}$ superconductor \cite{Rasolt}, and so
on. Further investigating the applications of the results to the
high $T_c$ superconductivity, the state preparation of the
harmonically trapped ions and the quantum computation will be
interesting.

\begin{acknowledgments}
This work was supported by the National Natural Science Foundation
of China under Grant No. 10275023.
\end{acknowledgments}

\end{document}